\documentclass[12pt]{article}
\usepackage{times}
\usepackage{geometry}
\usepackage[utf8]{inputenc}
\usepackage{enumitem,amssymb}
\usepackage{ragged2e}
\newlist{thematic}{itemize}{8}
\setlist[thematic]{label=$\square$}
\usepackage{pifont}
\usepackage{natbib}
\usepackage{aastex_hack}
\usepackage{graphicx}
\usepackage{titlesec}
\titlespacing*{\section}{0pt}{12pt}{6pt}
\titleformat*{\section}{\normalsize\bfseries}
\titlespacing*{\subsection}{0pt}{12pt}{6pt}
\titleformat*{\subsection}{\normalsize\bfseries}

\newcommand{\JWST}{\textit{JWST}}
\newcommand{\HST}{\textit{HST}}
\newcommand{\Kepler}{\textit{Kepler}}
\newcommand{\TESS}{\textit{TESS}}

\begin{document}
\raggedright
\huge
Astro2020 Science White Paper \linebreak

Constraining Stellar Photospheres as an Essential Step for Transmission Spectroscopy of Small Exoplanets \linebreak
\normalsize

\noindent \textbf{Thematic Areas:} \hspace*{57pt}
$\boxtimes$ Planetary Systems \hspace*{13pt} 
$\square$ Star and Planet Formation \hspace*{20pt}\linebreak
$\square$ Formation and Evolution of Compact Objects \hspace*{31pt} 
$\square$ Cosmology and Fundamental Physics \linebreak
$\boxtimes$ Stars and Stellar Evolution \hspace*{1pt} 
$\square$ Resolved Stellar Populations and their Environments \hspace*{40pt} \linebreak
$\square$ Galaxy Evolution   \hspace*{45pt}
$\square$ Multi-Messenger Astronomy and Astrophysics \hspace*{65pt} \linebreak
  
\textbf{Principal Author:}

Name:	Benjamin V. Rackham
 \linebreak						
Institution:  Steward Observatory, University of Arizona
 \linebreak
Email:        brackham@as.arizona.edu
 \linebreak
 \linebreak
 
\textbf{Co-authors:} 
    
    Arazi Pinhas (University of Cambridge),

    D\'aniel Apai (University of Arizona),

    Rapha\"elle Haywood (Harvard-Smithsonian Center for Astrophysics),

    Heather Cegla (Geneva Observatory),

    N\'estor Espinoza (Max Planck Institute for Astronomy),

    Johanna K. Teske (Carnegie Observatories),

    Michael Gully-Santiago (NASA Ames Research Center),

    Gioia Rau (NASA Goddard Space Flight Center),

    Brett M. Morris (University of Washington),

    Daniel Angerhausen (University of Bern),

    Thomas Barclay (NASA Goddard Space Flight Center),

    Ludmila Carone (Max Planck Institute for Astronomy),

    P. Wilson Cauley (University of Colorado Boulder),

    Julien de Wit (Massachusetts Institute of Technology),

    Shawn Domagal-Goldman (NASA Goddard Space Flight Center),

    Chuanfei Dong (Princeton University),

    Diana Dragomir (Massachusetts Institute of Technology / University of New Mexico),
    
    Mark S. Giampapa (National Solar Observatory),
    
    Yasuhiro Hasegawa (Jet Propulsion Laboratory),
    
    Natalie R. Hinkel (Southwest Research Institute),
    
    Renyu Hu (Jet Propulsion Laboratory),
    
    Andr\'es Jord\'an (Pontifical Catholic University of Chile),
    
    Irina Kitiashvili (NASA Ames Research Center),
    
    Laura Kreidberg (Harvard-Smithsonian Center for Astrophysics),
    
    Carey Lisse (Johns Hopkins University),
    
    Joe Llama (Lowell Observatory),
    
    Mercedes L\'opez-Morales (Harvard-Smithsonian Center for Astrophysics),
    
    Bertrand Mennesson (Jet Propulsion Laboratory),
    
    Karan Molaverdikhani (Max Planck Institute for Astronomy),
    
    David J. Osip (Las Campanas Observatory),
    
    Elisa V. Quintana (NASA Goddard Space Flight Center)
  \linebreak

\textbf{Abstract:}
\justifying
Transiting exoplanets offer a unique opportunity to study the atmospheres of terrestrial worlds in other systems in the coming decade.
By absorbing and scattering starlight, exoplanet atmospheres produce spectroscopic transit depth variations that allow us to probe their physical structures and chemical compositions.
These same variations, however, can be introduced by the photospheric heterogeneity of the host star (i.e., the transit light source effect).
Recent modeling efforts and increasingly precise observations are revealing that our understanding of transmission spectra of the smallest transiting exoplanets will likely be limited by our knowledge of host star photospheres.

Here we outline promising scientific opportunities for the next decade that can provide useful constraints on stellar photospheres and inform interpretations of transmission spectra of the smallest ($R<4\,R_{\odot}$) exoplanets.
We identify and discuss four primary opportunities:
(1) refining stellar magnetic active region properties through exoplanet crossing events;
(2) spectral decomposition of active exoplanet host stars;
(3) joint retrievals of stellar photospheric and planetary atmospheric properties with studies of transmission spectra; and
(4) continued visual transmission spectroscopy studies to complement longer-wavelength studies from \JWST{}.

In this context, we make five recommendations to the Astro2020 Decadal Survey Committee:
(1) identify the transit light source (TLS) effect as a challenge to precise exoplanet transmission spectroscopy and an opportunity ripe for scientific advancement in the coming decade;
(2) include characterization of host star photospheric heterogeneity as part of a comprehensive research strategy for studying transiting exoplanets;
(3) support the construction of ground-based extremely large telescopes (ELTs);
(4) support multi-disciplinary research teams that bring together the heliophysics, stellar physics, and exoplanet communities to further exploit transiting exoplanets as spatial probes of stellar photospheres; and
(5) support visual transmission spectroscopy efforts as complements to longer-wavelength observational campaigns with \JWST{}.

\pagebreak

\section{The Transit Light Source Effect}

The last two decades have witnessed an explosion of theoretical and observational advances in the study of exoplanet atmospheres.
In large part, these advances have been enabled by exoplanet transmission spectroscopy, the multiwavelength study of exoplanet transit depths
This technique probes the thin upper atmospheres of distant worlds and provides unprecedented insights into their physical structures and chemical compositions.
Observations of transiting exoplanets have characterized atomic and molecular absorption \citep[e.g.,][]{Charbonneau2002, Deming2013}, scattering processes in upper atmospheres \citep[e.g.,][]{Lecavelier_Des_Etangs2008, Pont2013}, and the importance of clouds and hazes in shaping transmission spectra \citep[e.g.,][]{Kreidberg2014, Sing2016}.
The next decade promises even greater advancements in the characterization of giant exoplanets and a revolution in studies of terrestrial exoplanet atmospheres.
\TESS{} and focused ground-based transit surveys, such as MEarth \citep{Nutzman2008}, SPECULOOS \citep{SPECULOOS}, and Project EDEN\footnote{project-eden.space}, will discover small planets amenable for atmospheric characterization; 
\JWST{} and ground-based extremely large telescopes (ELTs) will deliver the collecting area, spectral coverage, and spectral resolution required to detect biomarkers through transmission spectroscopy \citep{Cowan2015, Schwieterman2018}.
If life beyond Earth is ubiquitous, we will---for the first time in human history---be able to potentially detect it \citep{Seager2014}, provided we can disentangle exoplanet atmospheric signatures from the intrinsic ``noise'' of their host stars.

\begin{figure}[tb]
    \centering
    \includegraphics[width=0.9\linewidth]{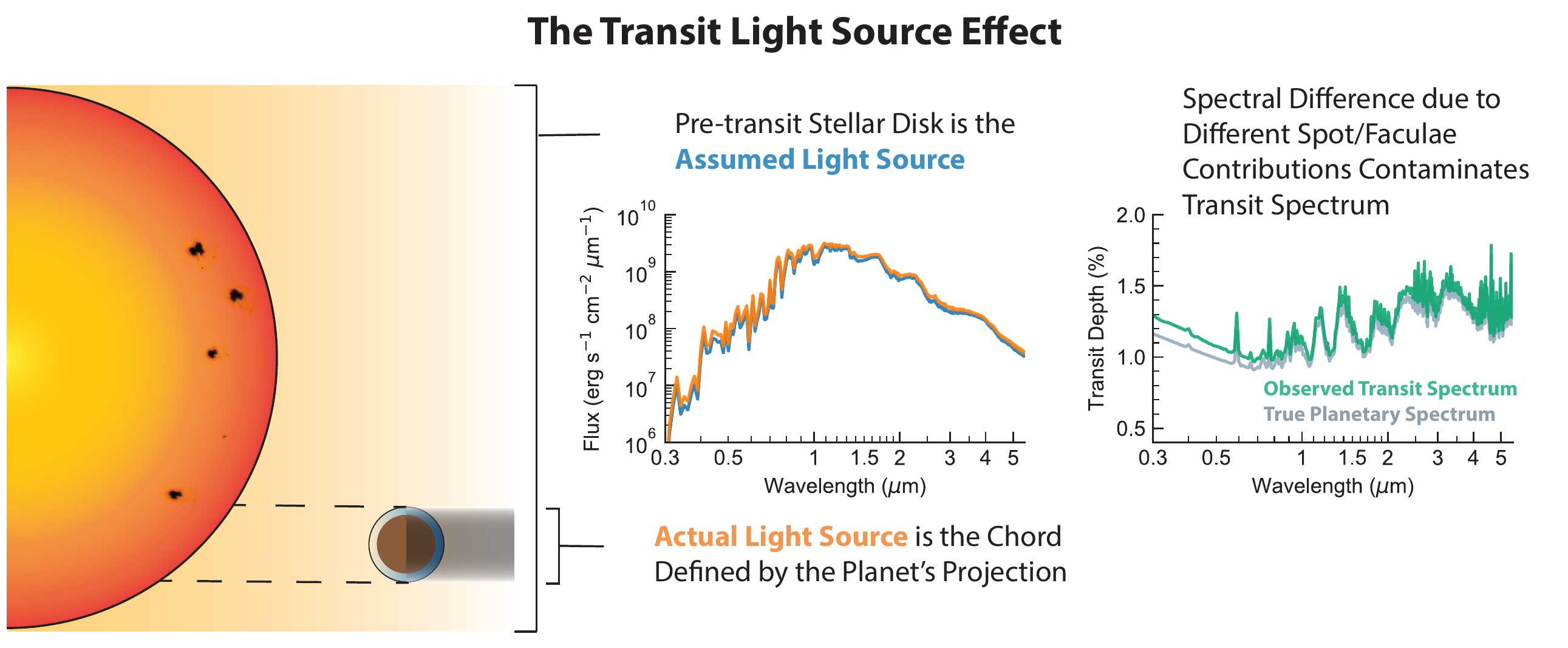}
    \caption{
Schematic of the transit light source effect.
Inhomogeneities in a stellar photosphere introduce a spectral difference between the light that illuminates the exoplanet atmosphere during a transit and the disk-integrated stellar spectrum, which provides the reference for measuring transit depths.
This spectral mismatch produces apparent transit depth variations that can mimic or mask exoplanetary atmospheric features.
From \citet{Rackham2018}.
    }
    \label{fig:TLSE}
\end{figure}

This caveat owes to recent modeling efforts and increasingly precise observations that have underscored the impact of stellar photospheric heterogeneity on precise transmission spectra.
In particular, stellar magnetic active regions in the form of cool spots and hot faculae, present both within and outside the transit chord, introduce strong signals that can alter the observed transmission spectrum of an exoplanet \citep[e.g.,][]{Sing2011, Pont2013, Oshagh2014, McCullough2014, Rackham2017}.
Since we cannot directly measure the emergent spectrum of the spatially resolved photospheric region that illuminates an exoplanet atmosphere during a transit, we must instead adopt the spectrum of the out-of-transit stellar disk as our reference.
Any difference between the disk-averaged spectrum and the spectrum of the transit chord---the actual light source for the measurement---will be imprinted on the transmission spectrum that we observe.
The importance of this phenomenon, known as the transit light source (TLS) effect \citep[][\textbf{Figure~\ref{fig:TLSE}}]{Rackham2018, Rackham2019}, was emphasized by the National Academies of Science, Engineering, and Medicine in its Exoplanet Science Strategy Consensus Study Report \citep{ESS}, which found that ”[u]nderstanding of exoplanets is limited by measurements of the properties of the parent stars, including [their]...emergent spectrum and variability” (p. S-5). 

A detailed discussion of the TLS effect, its astrophysical origin, and its impact on exoplanet characterization has been provided by \citet{Apai2018}.
That analysis identified three key questions that will be essential to address in order to develop a robust method of correcting transmission spectra for the TLS effect, which we paraphrase as:
(1) \textit{How do starspot and facula properties (size distribution, temperature distribution, and spatial distribution) vary with spectral type and stellar activity level?}
(2) \textit{What model components are required to describe TLS spectral signals due to stellar heterogeneity?}
(3) \textit{What observations are required by stellar heterogeneity models to calculate and predict TLS spectral signals for a given epoch?}
This white paper focuses on the science opportunities in the coming decade to address these questions and mitigate the impact of stellar photospheric heterogeneity on exoplanet transmission spectroscopy.
In the broader context of host star characterization, this white paper complements separate, complementary papers on outstanding questions and opportunities for host star characterization generally \citep{Hinkel2019a} and stellar abundances \citep{Hinkel2019b}.

\section{Science Opportunities}

While certainly not an exhaustive list, we identify four promising observational and data-analysis techniques that can be further developed or exploited in the next ten years to advance our understanding of the photospheres of exoplanet host stars.
These are detailed in the following sections.

\subsection{Refining active region properties through exoplanet crossing events}

\begin{figure}[tb]
    \centering
    \includegraphics[height=0.43\textheight]{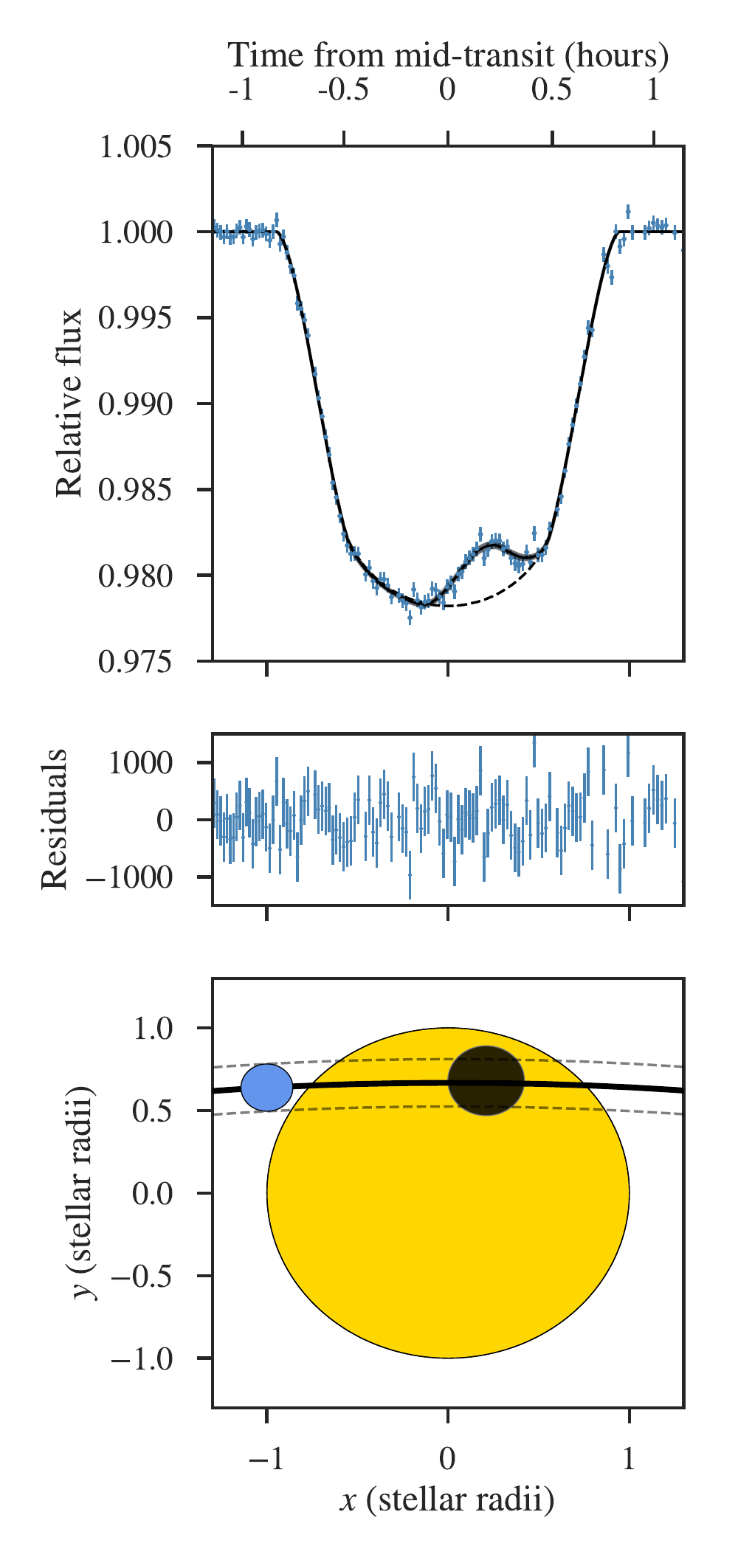}
    \includegraphics[height=0.43\textheight]{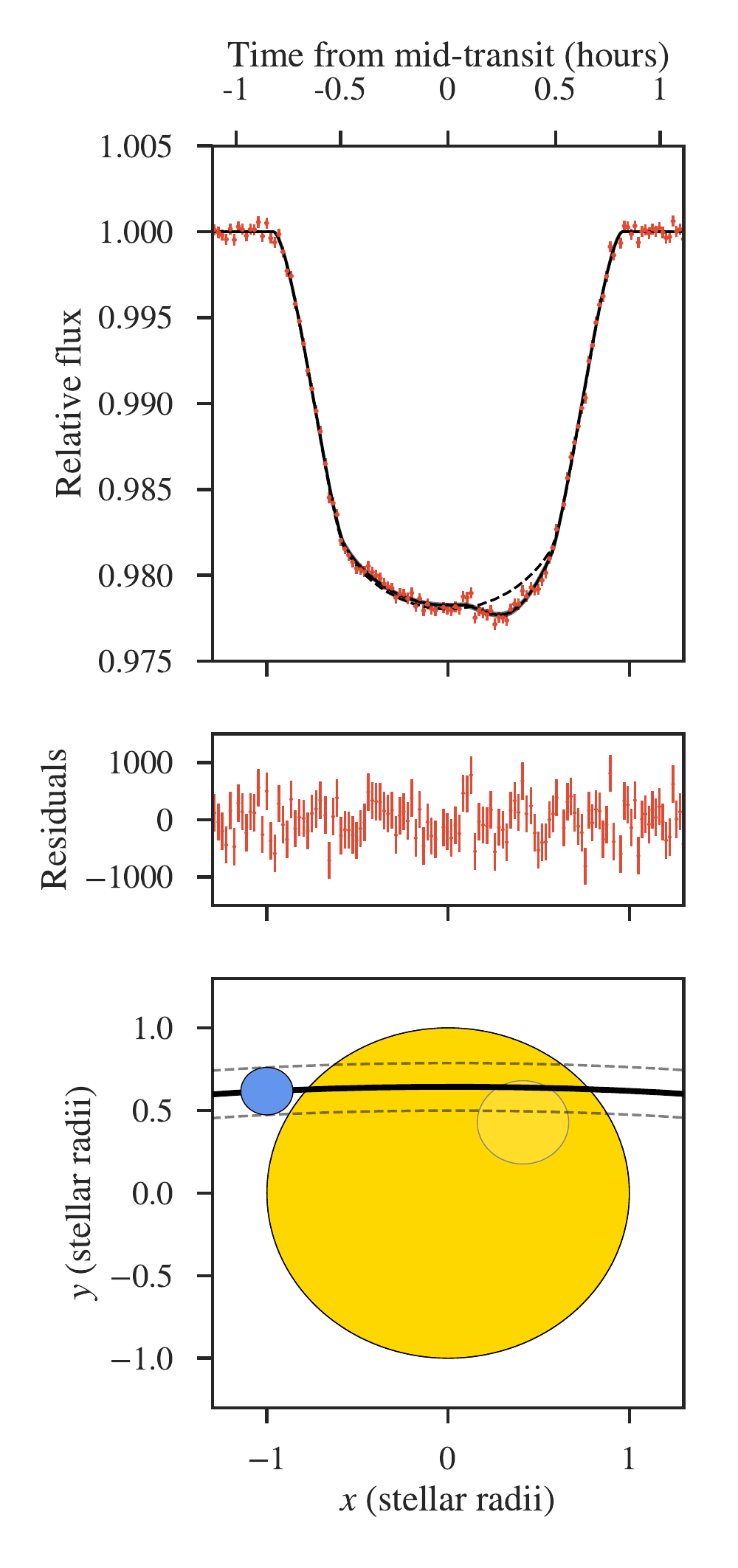}
    \includegraphics[height=0.43\textheight]{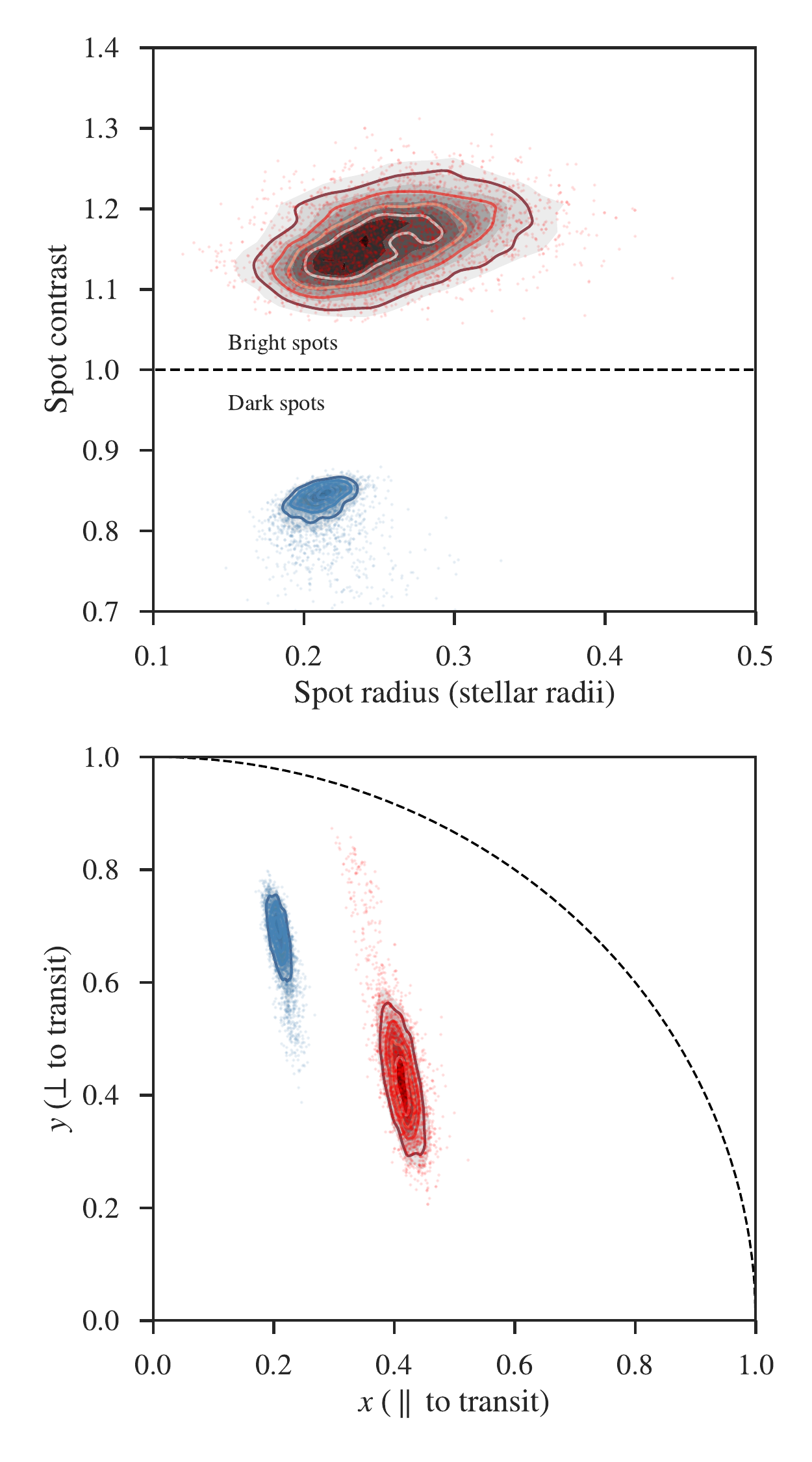}
    \caption{
Identification of spot (left panels) and facula (middle panels) properties through active region crossing events.
Precise transit light curves constrain the sizes, contrasts, and positions of cool (spotted) and hot (facular) magnetic active regions in stellar photospheres (right panels).
Spectroscopic transit observations have the added benefit of studying the wavelength dependence of the active region contrast, which can be used to constrain the temperature contrast of the active region with respect to the immaculate photosphere.
From \citet{Espinoza2019}.
    }
    \label{fig:Espinoza_2019}
\end{figure}

Attempts to constrain the TLS effect are limited by our incomplete understanding of the properties of stellar magnetic active regions.
In particular, uncertainties in the typical sizes, emergent spectra, and spatial distributions of active regions drive the uncertainties in forward-modeling approaches to forecast TLS signals for typical exoplanet host stars \citep{Rackham2018, Rackham2019}.
Fortunately, transiting exoplanets provide a spatial probe of stellar photospheres that can be leveraged to extract these properties.
Transiting exoplanets may occult magnetic active regions in the stellar photosphere, producing notable deviations in light curves from models of transits of uniform photospheres (\textbf{Figure~\ref{fig:Espinoza_2019}}).
The timing, duration, and amplitude of these active region crossing events respectively encode the position, size, and contrast of active regions in the stellar photosphere.
In the past decade, this technique has enabled studies of the magnetic active regions in exemplar systems, such as HD~189733 \citep{Sing2011, Pont2013}, HAT-P-11 \citep{SPOTROD, Morris2017}, and WASP-19 \citep{Mancini2013, Espinoza2019}.
In the next decade, future work can use the large datasets provided by \Kepler{} and \TESS{} and new analysis tools, such as \texttt{SPOTROD} \citep{SPOTROD}, \texttt{StarSim} \citep{StarSim}, and \texttt{PyTransSpot} \citep{PyTranSpot}, to explore the dependence of active region properties on spectral type, age, activity level, and other variables of the star.
Incorporating 3D MHD codes, such as \texttt{MURaM}, into these analyses \citep[e.g.,][]{Norris2017}, can help to further reveal stellar physics from transit observations.

\subsection{Spectral decomposition of active exoplanet host stars}

\begin{figure}[tb]
    \centering
    \includegraphics[width=0.9\linewidth]{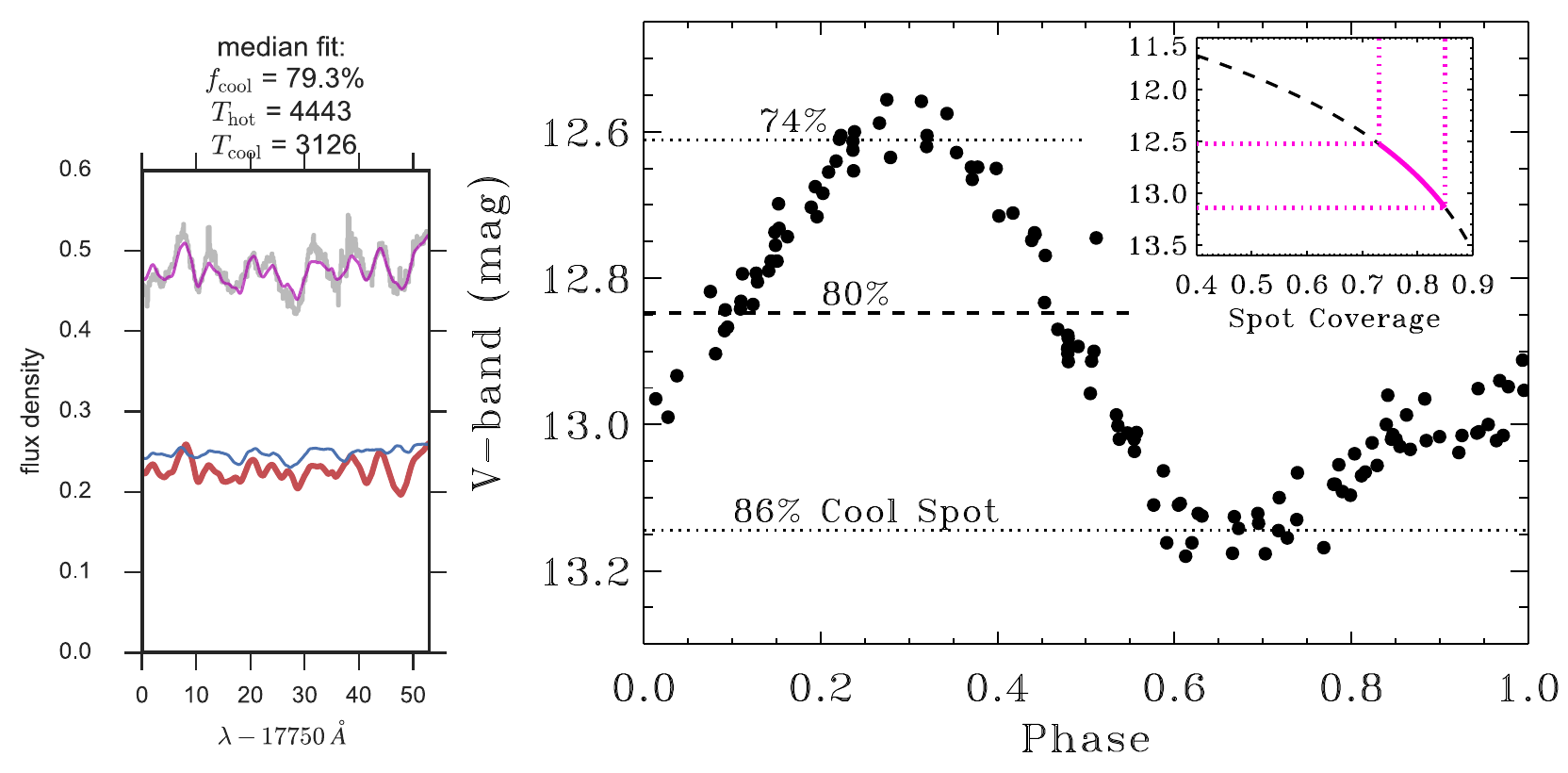}
    \caption{
Illustration of the spectral decomposition technique. Features in the high-resolution NIR spectrum of LkCa~4 reveal the temperatures and covering fractions of a cool and hot photospheric component at a single epoch (left). Combined with long-term photometric monitoring (right), this information can be used to determine the covering fraction of the cool component as the star rotates.
Adapted from \citet{Gully-Santiago2017}.}
    \label{fig:spec_decomp}
\end{figure}

Spectral decomposition provides another avenue to explore the heterogeneity of stellar photospheres.
This technique involves fitting features in stellar spectra that must arise in different temperature regimes to constrain the temperatures and covering fractions of the unique spectral components in an unresolved stellar photosphere \citep{Vogt1979, Ramsey1980, Vogt1981}.
For example, TiO bands apparent in the medium-resolution ($R {\sim} 10,000$) visual spectra of active G and K stars point to starspot covering fractions as large as 64\% in some cases \citep{Neff1995, ONeal1996, ONeal1998}.
Similarly, TiO band strengths in $R {\sim} 1800$ spectra of 304 active GKM dwarfs that are candidate Pleiades members indicate starspot covering fractions (${\sim} 50\%$) for many K and M stars, even those with little or no brightness variations \citep{Fang2016}.

Recent work has shown the strength of new robust spectral inference frameworks
for carrying out this work over a wider wavelength range and exploiting model stellar spectra grids to explore stellar heterogeneity over a wide parameter space.
In particular, \citet{Gully-Santiago2017} studied the T~Tauri star LkCa~4 using a suite of visual and near-infrared (NIR) observations and found clear evidence for a hotter (${\sim} 4100 \, \rm K$) and cooler (${\sim} 2700$--$3000 \, \rm K$) photospheric component.
High-resolution NIR spectra suggest a covering fraction of ${\sim} 80\%$ for the cooler component, the rotational modulation of which can be traced with long-term photometric monitoring \citep[][\textbf{Figure~\ref{fig:spec_decomp}}]{Gully-Santiago2017}.
The combination of spectral decomposition and long-term photometric monitoring offers a powerful approach for constraining the time-resolved heterogeneity of stellar photospheres, accessing both persistent and rotationally modulated active regions.

The potential of studying exoplanet host stars with this technique is clear.
The low-resolution ($R {\sim} 130$) NIR spectrum of terrestrial exoplanet host TRAPPIST-1, for example, shows evidence for three spectral components in its photosphere, including a large (potentially dominant) cool component \citep{Zhang2018, Wakeford2019}.
The photospheric heterogeneity of this ultracool dwarf may produce TLS signals in the transmission spectra of the TRAPPIST-1 planets that are comparable to planetary signals \citep[e.g.,][]{Ducrot2018} or even an order of magnitude greater than them \citep[e.g.,][]{Zhang2018} at visual and NIR wavelengths.
In the next decade, future work should exploit recent advances in spectral decomposition techniques to study the time-resolved heterogeneity of TRAPPIST-1 and other high-priority active exoplanet host stars.
Modeling efforts should also explore wavelength ranges and spectral resolutions necessary for constraints on active regions that will be useful for precise exoplanet transmission spectroscopy.

\subsection{Joint retrievals of stellar and planetary properties}

Retrieval methodologies that have traditionally been used to infer planetary atmospheric properties from transmission spectra offer a promising approach for inferring stellar properties from transmission spectra as well.
The contribution of unocculted active regions to observed transmission spectra via the TLS effect can be modeled simply in retrievals with a few additional parameters, such as the temperature contrast of the active region with the immaculate photosphere and the active region covering fraction \citep{Pinhas2018, Espinoza2019}.
In a nested sampling framework \citep{Skilling2006}, the Bayesian evidence for models with and without TLS signals can be compared straightforwardly to assess whether the data warrant the additional complexity of a heterogeneous stellar photosphere.
Recent studies employing this retrieval framework have found examples of transmission spectra affected by unocculted spots \citep[e.g., WASP-19b;][]{Espinoza2019} and others unaffected by starspots beyond those occulted during transits \citep[e.g., WASP-4b;][]{Bixel2019}.
Applied to a comparative sample of nine high-quality hot Jupiter transmission spectra \citep{Sing2016}, retrievals allowing for TLS signals indicate evidence for stellar heterogeneity impacting a third of these spectra \citep{Pinhas2018}.
These studies illustrate the strength and future potential of this approach, which should be developed further in the coming decade.
They also highlight the need for informed priors on the parameters of stellar active regions (e.g., temperatures and covering fractions), which can be provided in the coming decade by some of the strategies that we espouse here.

\subsection{Visual transmission spectroscopy in the era of \JWST{}}

Stellar heterogeneity has a larger impact on transit observations at shorter wavelengths, owing to the larger contrasts between active regions and the immaculate photosphere.
Unocculted spots, for example, cause transit depths to increase toward shorter wavelengths, while unocculted faculae decrease them \citep[e.g.,][]{Pont2008, Oshagh2014}.
In specific temperature regimes, opacity differences between active regions and the immaculate photosphere can even imprint molecular features, such as those of TiO, on transmission spectra \citep{Espinoza2019}.
This fact complicates interpretations of visual transmission spectra alone, but it also provides an opportunity: in a joint retrieval framework, visual spectra can constrain TLS signals (or the lack thereof) that are subtler are longer wavelengths \citep[e.g.,][]{Rackham2017}.
In the coming era of \JWST{}, visual transmission spectra will provide crucial complements to \JWST{} observations at longer wavelengths.
Observational efforts to measure visual transmission spectra with \HST{} and ground-based facilities will remain vital to the interpretation of longer-wavelength data.

\section{Recommendations}

We make the following recommendations to the Astro2020 Decadal Survey Committee:
(1) Identify the TLS effect as a challenge to precise exoplanet transmission spectroscopy and an opportunity ripe for scientific advancement in the coming decade.
(2) Include characterization of host star photospheric heterogeneity as part of a comprehensive research strategy for studying transiting exoplanets.
(3) Support the construction of ground-based ELTs. For each of the science opportunities outlined here, the light-gathering capacities of the ELTs will enable precise characterization of an unprecedented sample of exoplanet host stars.
(4) Support multi-disciplinary research teams that bring together the heliophysics, stellar physics, and exoplanet communities to further exploit transiting exoplanets as spatial probes of stellar photospheres.
(5) Support visual transmission spectroscopy efforts as complements to longer-wavelength observational campaigns with \JWST{}.

\pagebreak
\bibliographystyle{aasjournal}
\bibliography{references}

\end{document}